\documentstyle[preprint,aps,eqsecnum]{revtex}
\begin{document}

\begin{center}
  {\Large \bf Manifolds in random media: A variational
    approach to the spatial probability distribution} \vspace{.2in}

by\\

\vspace{.15in}

Yadin Y. Goldschmidt\\

Department of Physics and Astronomy\\
University of Pittsburgh\\
Pittsburgh, PA  15260
\end{center}

\vspace{.15in}

\begin{center}
  {\bf {Abstract}}
\end{center}
\vspace{0.15in}

We develop a new variational scheme to approximate the
position dependent spatial probability distribution
of a zero dimensional manifold in a random medium.
This celebrated 'toy-model' is associated via a mapping with directed
polymers in 1+1 dimension, and also describes features of the
commensurate-incommensurate phase transition. It consists of a pointlike
'interface' in one dimension subject to a combination of a
harmonic potential plus a random potential with long range spatial
correlations. The variational approach we develop gives far better
results for the tail of the spatial distribution than the hamiltonian
version, developed by Mezard and Parisi, as compared with numerical
simulations for a range of temperatures. This is because the
variational parameters are determined as functions of position.
The replica method is utilized, and solutions for the variational
parameters are presented. In this paper we limit ourselves
to the replica symmetric solution.

\vspace{0.2in}


\newpage

\section{{\bf Introduction}}
Recently a lot of attention has been devoted to the behavior of
manifolds in random media.
\cite{villain,2,3,brezin,parisi,orland,7,mp1,halpin,10,bmy,12,mp,zumophen,gb1,gb2,yg1,engel,yg2,blum,ledoussal}
This is partially due to their connection with
vortex line pinning in high $T_c$ superconductors, \cite{bmy,ledoussal}
but also because of
the intrinsic interest in the behavior of interfaces between two
coexisting phases of a disordered system, like magnets subject to
random fields or random impurities. In addition mappings are known to
exist between one dimensional manifolds like directed polymers in
disordered media and growth problems in the presence of random noise
described by the KPZ equation.\cite{kpz}

The variational method \cite{mp1,mp,gb2}
appears to be an important tool in approximating many properties
of the system like the calculation of the roughness exponent
of a wandering manifold in a disordered medium. Recently we
have conducted a numerical study of the spatial probability distribution
of directed polymers in 1+1 dimensions in the presence of quenched
disorder. \cite{gb1} Directed polymers refer to an interface in two
dimensions with no overhangs. In addition to the width of the
interface, it is important to know what is the probability that the
interface would wander a certain distance in the transverse direction.
We have then exploited a mapping
of this system into that of a pointlike interface in one dimension
subjected to a combination of a harmonic restoring force and a random
potential with long ranged correlations.

Using that mapping combined
with a use of the variational approximation, we were able to derive many
properties of the spatial probability distribution (PD) of the directed
polymers. \cite{gb2} In particular the function f($\alpha$) associated
with the decay
of the probability for different realizations of the disorder has been
derived using the replica method. Replica symmetry breaking solutions
have been used (at low temperatures). The mapping from directed
polymers to the so called
toy-model is discussed in detail in ref. \cite{gb2} based on previous
work by Parisi \cite{parisi} and Bouchaud and Orland \cite{orland}.
In Particular the spatial probability distribution of directed polymers
at large times relates to the spatial probability distribution of
the toy-model at low temperature.
Villain {\it et al.} \cite{villain}
emphasize directly the relation between the toy-model
and the commensurate-incommensurate transition in systems with
quenched impurities. \cite{villain}

Using the replica method the toy-model can be mapped into an $n$-body
problem were the interaction between different particles (replicas)
is given by the correlation of the original random potential. The
variational method utilizes an effective quadratic hamiltonian
whose parameters are chosen by the stationarity method to give a good
approximation for the free energy of the $n$-body system in the limit $n
\rightarrow 0$. \cite{mp1}
The results it provides for the tail of the spatial distribution are
not very accurate as will be shown below (Section 4.).
In this paper we describe a way to directly apply
the variational approach to the probability distribution rather than
to the hamiltonian or the free energy. As we shall see it gives a much
better approximation for the behavior of the tail of the
distribution. For high temperatures the results
appear almost exact without the need to break replica symmetry.

\section{{\bf The variational method applied to the probability
  distribution.}}
The toy model \cite{villain,brezin,mp,gb2} involves a classical
particle in a
one-dimensional potential consisting of a harmonic part and a
spatially correlated random part. It is the simple prototype of an
interface-a point in one dimension- which is localized by a harmonic
restoring force, and pinned by quenched impurities the distribution of
which may be correlated over large separations.
\ Its hamiltonian is given by:
\begin{equation}
H (\omega)\, = \, \frac{\mu}{2} \, \omega^2 + V(\omega) ,
\end{equation}
where $\omega$ denotes the position of the particle and
$V(\omega)$ is a random potential with a gaussian distribution
characterized by:
\begin{equation}
\langle V(\omega)\rangle \, =\, 0, \, \, \, \, \, \langle V(\omega)
V(\omega^{\prime})\rangle \, = \, -\frac{g}{2(1-\gamma)}\,
|\omega-\omega^{\prime}|^{2- 2 \gamma} \, + \, {\rm Const.} \equiv -f((\omega
- \omega')^2),
\label{eq:f}
\end{equation}
where the brackets indicate averaging over the random potential.
In the sequel without loss of generality we will take the constant on
the r.h.s. of eq.(\ref{eq:f}) to be zero.
We will make use of the replica method to replicate the partition
function
\begin{equation}
Z\ = \ \int d \omega \ \exp (\ - \beta H \ )
  \label{eq:Z}
\end{equation}
and average it over the random
potential to express it in terms of an effective $n$-body hamiltonian:
\begin{eqnarray}
\langle Z^n \rangle \ =\ \int d\omega_1 \cdots d\omega_n \ \exp (-\beta
{\cal H})\ , \hspace{.5in} \nonumber \\
{\cal H} \ = \ {1 \over 2} \mu \sum_{a=1}^n \omega_a^2 \ + \ {1 \over
  2} \beta \sum_{ab} \ f\ (\ (\omega_a\ - \ \omega_b\ )^2\ )\ .
  \label{eq:H}
\end{eqnarray}

For later use let us also define an $n$-body quadratic hamiltonian:
\begin{equation}
h\ = \ {1 \over 2} \mu \sum_{a=1}^n \omega_a^2 \ - \ {1 \over
  2} \beta \sum_{ab} \sigma_{ab}\ \omega_a \ \omega_b
  \label{eq:h}
\end{equation}
where the matrix elements of $\sigma_{ab}$ are free
parameters at this point.

We consider cases in which the potential has long-range
correlations, ({\em i.e.} $\gamma<1$), and in particular we are
interested in the case $\gamma=\frac{1}{2}$ because of the above mentioned
mapping from directed polymers in $1+1$ dimensions.
Let us review briefly the variational scheme used by Mezard and Parisi.
We start with the well known inequality
\begin{equation}
  \langle e^A \rangle \ \ \ \geq \ \ \ e^{\ \langle A \rangle}
  \label{eq:ineq}
\end{equation}
The Mezard-Parisi variational approximation \cite{mp1} can be
obtained from this inequality, by choosing
\begin{equation}
A(\omega_1, \cdots ,\omega_n)={\cal H}(\omega_1, \cdots ,\omega_n) -
h(\omega_1, \cdots ,\omega_n)
  \label{eq:A}
\end{equation}
And by defining the average as
\begin{equation}
\langle A \rangle \ =\ {Tr A \ \exp \ (\ -\beta h\ ) \over Tr \exp \
  (\ -\beta h)}.
  \label{eq:av1}
\end{equation}
with
\begin{equation}
Tr \ =\ \int_{-\infty}^{+\infty} d\omega_1\ \cdots\ d\omega_n
  \label{eq:tr}
\end{equation}
This yields
\begin{equation}
n f\ =\ -{1 \over \beta} \ln Tr \exp \ (\ -\beta {\cal H}\ )\ \leq \
\langle{\cal H} - h\ \rangle \ -\ {1 \over \beta}\ Tr \ \ln \ \exp \
(-\beta h\ ),
  \label{eq:fe}
\end{equation}
where f is the free energy.
The variational free energy is given by the right hand side of eq.
(\ref{eq:fe}) at the point of stationarity. This procedure provides
equations for the variational parameters which are the elements of the
matrix $\sigma_{ab}$.
MP \cite{mp} have solved these equations and found a replica symmetric
(RS) solution
valid for high temperatures and a solution which breaks replica
symmetry (RSB) at low temperatures.

Recall that the limit $n \rightarrow 0$ has to be taken when making
use of the replica method. An important point to bear in mind is that the
inequality (\ref{eq:ineq}) holds only when $n$ is a positive integer,
provided the parametrization of $\sigma$ is such that $Tr\ \exp(-\beta
h) < \infty$. This is because only in this case the integration measure
\begin{equation}
[\ Tr \exp (-\beta h)]^{-1} \exp\ (-\beta h) \ d\omega_1 \cdots d\omega_n ,
  \label{eq:meas1}
\end{equation}
is a real positive measure on $R^n$, for which the proof
of the inequality (\ref{eq:ineq}) holds (see e.g. \cite{rudin} p. 61).
In the limit $n \rightarrow 0$ the inequality can change sign
and does not hold in general, but one still expects the the stationary
point of the r.h.s. of equation (\ref{eq:fe}) to give a good
approximation to the l.h.s. i.e. to the exact free energy. Support for
this contention comes also from the fact that the variational method
becomes exact for a manifold embedded in $d$ spatial dimensions when
$d$ becomes infinite, and one can systematically improve it by a $1/d$
expansion. \cite{mp1,mp,yg1,yg2}

Let us define
\begin{equation}
\widehat{Tr}\ = \ \int d\omega_1 \cdots d\omega_n \ \delta ( \omega_1\
- \ \omega ) \ .
  \label{eq:trhat}
\end{equation}
The function
\begin{equation}
P_h\ (\omega) \ = \ \widehat{Tr} \ \exp\ (-\beta h\ ) \ ,
  \label{eq:Ph}
\end{equation}
(with the parameters $\sigma_{\alpha\beta}$ determined by the
stationarity conditions which were discussed above)
constitutes an approximation to the exact spatial probability
distribution, averaged over the random realizations of the potential,
and  given by the formula
\begin{equation}
P_H\ (\omega) \equiv \left\langle {\exp (-\beta H (\omega)) \over \int
d\sigma \exp (-\beta H(\sigma))} \right\rangle \ = \ \widehat{Tr} \
\exp\ (-\beta {\cal H}\ ) \ ,
 \label{eq:PH}
\end{equation}
with the limit $n \rightarrow 0$ to be understood.
Eq. (\ref{eq:Ph}) has been evaluated by us in a previous publication
\cite{gb2} and compared with numerical simulations both using the RS
and RSB solutions for the variational parameters. The result for $P_h$
is \cite{gb2}:
\begin{equation}
P_h\ ( \omega\ ) = \ \left({\beta \over 2 \pi G_{11}}\right)^{1/2}
\exp \left({-\beta \omega^2 \over
  2 G_{11}}\right)
  \label{eq:Phf}
\end{equation}
with
\begin{equation}
G_{ab}\ = \ [\ \mu I \ -\ \hat{\sigma}\ ]^{-1}_{ab}.
  \label{eq:G}
\end{equation}
Substituting the RS and RSB solutions for $G_{11}$ obtained in ref.
(\cite{mp}), we find \cite{gb2}
\begin{eqnarray}
P_h\ (\ \omega\ ) = \ \left({\beta \mu \over 2 \pi (1+\gamma^{-1}
\hat{T}^{-(1+\gamma)})}\right)^{1/2}  \exp\left(-{\beta \mu \omega^2
\over 2(1+\gamma^{-1} \ \hat{T}^{-(1+\gamma)})}\right), \hspace{.2in}
\hat{T} > 1 \nonumber \\
P_h\ (\ \omega\ ) = \ \left( {\beta \mu \gamma \hat{T} \over 2\pi(1 +\gamma)}
\right)^{1/2} \exp \left( -{\beta \mu \gamma \hat{T} \over 2 (1 +\gamma)}
\ \omega^2 \right), \hspace{1.4in} \hat{T}<1
\label{eq:Phexp}
\end{eqnarray}
with the 'reduced' temperature $\hat{T}$ defined as
\begin{equation}
\hat{T}\ =\ \beta^{-1} \mu^{{1-\gamma \over 1+\gamma}} \left(\gamma 2^{1-2
\gamma} {\Gamma(3/2-\gamma) \over \Gamma(1/2)} g\right)^{-{1 \over
  1+\gamma}}.
\end{equation}
In the sequel we refer to the expressions given in eq.(\ref{eq:Phexp})
as ``the hamiltonian variational approximation''.

I proceed to derive a new variational scheme which is more
appropriate for approximating the position-dependent spatial
probability distribution. The idea is to use the
inequality (\ref{eq:ineq}), but substitute $\widehat{Tr}$ from eq.
(\ref{eq:trhat})
for the trace in eq. (\ref{eq:av1}). This means, defining
\begin{equation}
\langle \langle A \rangle \rangle (\omega)\ \equiv \ {\widehat{Tr}\ [
  A(\omega_1,\cdots,\omega_n) \ \exp \ (\ -\beta h\ )\ ] \over
  \widehat{Tr} \exp \
  (\ -\beta h\ )}.
  \label{eq:av2}
\end{equation}
The notation $\langle\langle\rangle\rangle$ is used to distinguish the
$\omega$- dependent average from the averaged defined in eq.(\ref{eq:av1}).
Using $A={\cal H} \ - \ h$ I find
\begin{equation}
P_H (\omega) = \ \widehat{Tr} \exp \left(\ \ -\beta {\cal H}) \ \geq
\ \exp \ (-\beta \langle \langle {\cal H} -h \rangle \rangle (\omega)\
\right)\ \times\
\widehat{Tr} \exp \ ( -\beta h)\ \equiv \ P_v(\omega)
  \label{eq:ineq2}
\end{equation}

Again, for positive integers $n$, and $\widehat{Tr} \exp \
  (\ -\beta h) < \infty$, the measure
\begin{equation}
[\ \widehat{Tr} \exp (-\beta h)]^{-1} \exp\ (-\beta h) \
\delta(\omega_1\ -\ \omega)\ d\omega_1 \cdots d\omega_n  \ \ \ ,
  \label{eq:meas2}
\end{equation}
is a positive measure on $R^n$ for any value of $\omega$ and the inequality
(\ref{eq:ineq}) holds. For $n \rightarrow 0$ this is no longer true,
but the stationary point of the r.h.s. of eq. (\ref{eq:ineq2}), denoted
by $P_v(\omega)$, with
respect to the parameters $\sigma_{ab}$
is expected to yield a good approximation to the spatial probability
distribution by virtue of analytic continuation. This will be checked
in Sec. 4
in comparison with numerical simulations. Such a variational
determination of $P_v(\omega)$ is obtained for each value of $\omega$
separately and thus it is expected to provide a better approximation
than by just using the $\sigma_{ab}$ obtained globally from extremizing the
free energy (see eq. (\ref{eq:fe})), and using eq. (\ref{eq:Ph}). One should
note that it does not follow from our discussion that $P_v(\omega)$ is
properly normalized. Provided it is a good approximation to the exact
probability distribution, its normalization will be close to one, and
correcting for the exact normalization will have practically no
effect on the behavior in the tail region were the probability is very small
and where the variational calculation is most important.

Our next task is to evaluate the quantity $\langle \langle {\cal H}-h\
\rangle \rangle (\omega)$ needed in order to calculate $P_v(\omega)$,
see eq. (\ref{eq:ineq2}).
If we use the integral representation for  the  Dirac
$\delta$-function in eq.(\ref{eq:trhat}):
\begin{equation}
\delta(\omega_1\  - \ \omega )\ =\ \int {dk \over 2 \pi} \
e^{ik\omega\ -\ ik\omega_1} \ \ .
\label{eq:delta}
\end{equation}
we can write

\begin{eqnarray}
\widehat{Tr} \ [({\cal  H}\ -\ h\ )\ e^{\ -\beta h}\ ] = \int {dk
  \over 2 \pi} e^{ik\omega}\
\int d\omega_1 \cdots d\omega_n  \ \ \ \ \ \ \ \ \  \ \ \ \ \ \ \ \ \
\ \ \ \ \ \ \ \ \ \ \ \ \ \ \ \ \ \ \  \nonumber  \\
\left({\beta \over 2}  \sum_{ab}
f((\omega_a-\omega_b)^2) + {1 \over 2} \sum_{ab}\sigma_{ab} \omega_a
\omega_b \right)
\times \exp \left( -{\beta   \over 2} \sum_{ab} (G^{-1})_{ab}
\omega_a\omega_b -ik\omega_1 \right)  \ .
\label{eq:trHmh}
\end{eqnarray}
In the Appendix we show how the integrals can be evaluated for general
$n$. The end result is:

\begin{eqnarray}
\langle \langle {\cal H}-h\ \rangle \rangle (\omega) = \hspace{4.8in}
\nonumber \\
{\beta \over 2}
\left( {G_{11} \over \pi}\right)^{1/2} \exp({\beta \omega^2 \over 2
    G_{11}}) \sum_{ab}\left[
\left(Z_{ab}\right)^{-1/2} \int_{-\infty}^{\infty} dp \ \exp(-p^2
 - {(\omega \sqrt{\beta}+p \sqrt{2}Y_{ab}   )^2 \over 2 Z_{ab} } )
\right. \hspace{0.6in}
\nonumber \\
\left.\times f\left({2 \over \beta} X_{ab} p^2 \right)\right]
- {n \over 2 \beta} + {\mu \over 2 \beta} \sum_{a}\ G_{aa}\ +\  {1
  \over 2 \beta G_{11}}
\ (G_{11} - \mu \sum_a G_{1a}^2) (1 - {\beta \omega^2 \over G_{11}}) \
\ \ \ \ \ \ \ \ \ \ \
\label{eq:avHmh}
\end{eqnarray}
with

\begin{eqnarray}
X_{ab}\ =\ G_{aa}\ +\ G_{bb}\ -\ 2 G_{ab}  \ \  \nonumber    \\
Y_{ab}\ =\ (G_{1a} - G_{1b}) \ / \ (X_{ab})^{1/2}  \nonumber \\
Z_{ab}\ =\ G_{11}\ -\ Y_{ab}^2 \ \ . \ \ \ \ \ \ \
\end{eqnarray}

For the special case of interest $\gamma = 1/2$ (see eq.(\ref{eq:f}) for
$f$) we find:
\begin{eqnarray}
\langle \langle {\cal H}\ -h \rangle \rangle (\omega) \ =\
\hspace{4.5in}\nonumber \\
g \sum_{ab}\left[ {\beta Y_{ab} X_{ab}^{1/2} \omega\over 2 G_{11}}\
{\rm erf} \left( \left(
\  {\beta Y_{ab}^2 \over 2 Z_{ab} G_{11}} \right)^{1/2}
\omega \right)
+ \left({\beta X_{ab} Z_{ab} \over 2 \pi G_{11}}\right)^{1/2}
\exp \left(-\ {\beta Y_{ab}^2 \omega^2 \over 2 Z_{ab} G_{11}}\
 \right) \ \right] \ \
\nonumber \\ - {n \over 2 \beta} + {\mu \over 2 \beta}
\sum_{a}\ G_{aa}\ +\  {1 \over 2 \beta G_{11}}
\ (G_{11} - \mu \sum_a G_{1a}^2)\ \ (1 - \beta \omega^2 / G_{11})\ ,
\hspace{1.2in}
\label{eq:avghalf}
\end{eqnarray}
where erf is the usual error function. Eqs. (\ref{eq:avHmh}) and
(\ref{eq:avghalf}) together with the expression (\ref{eq:ineq2}) for
$P_v(\omega)$ constitute the main results of this section.
In the rest of the paper we consider only the important case of $\gamma=1/2$.

\section{{\bf The replica symmetric solution}}
In this section we consider the replica symmetric solution to the
variational stationarity equations. In this case we search for a
solution for which all the off diagonal elements of the matrix
$\sigma_{ab}$ are equal as well as all the diagonal elements:
\begin{eqnarray}
\sigma_{ab}=\sigma  \ \ \ \ \ a \neq b \ \ \ \ \ \nonumber \\
\sigma_{aa}=\sigma_d  \ \ \ \ \ a=1 \cdots n
\label{eq:s}
\end{eqnarray}
we also define
\begin{eqnarray}
\mu_1=\mu+\sigma_d+(n-1)\sigma
\end{eqnarray}
In terms of these parameters one finds
\begin{eqnarray}
G_{aa}={1 \over \mu_1}\ \left( 1+{\sigma \over \mu_1} \right)\ \ ; \
\ \ G_{ab}= {\sigma \over \mu_1^2}\ ,\ \ \ \ a \neq b
\end{eqnarray}
Thus we have two variational parameters $\sigma$ and $\mu_1$ as
contrasted with the hamiltonian variational approach were there is
only one variable because in that case translational invariance
dictates $\mu_1=\mu$. In addition, the variational parameters are
of course dependent on $\omega$ in the present case.
We define for convenience:
\begin{eqnarray}
\Gamma =G_{11}+G_{12}.
\label{eq:Gamma}
\end{eqnarray}
Using the result of the previous section, eq.(\ref{eq:avghalf}),
we can express the probability distribution $P_v(\omega)$\ \
(see also eqs.(\ref{eq:ineq2}) and (\ref{eq:Ph})\ ) as
\begin{eqnarray}
P_v(\omega)= \left({\beta \over 2 \pi G_{11}}\right)^{1/2}
\exp \left(- {\beta \omega ^2 \over 2 G_{11}}\right) \
\exp\ (\ g(\omega )\ ) \hspace{1.in}\nonumber \\
g(\omega )= {g \beta ^2 \omega \over \mu_1 G_{11}}
\ {\rm erf}\left(\left({\beta \over 2 \mu_1 \Gamma
 G_{11}}\right)^{1/2} \omega \right)+g \beta \left(
{2 \beta \Gamma \over \pi \mu _1 G_{11}}\right)^{1/2}
\exp \left(-{\beta \omega ^2 \over 2 \mu_1 \Gamma G_{11}}\right)
\nonumber \\ - 2g \beta \left({\beta \over \pi \mu _1}\right)^{1/2}-
{1\over 2}\ \left(1-{\mu \Gamma \over \mu _1 G_{11}}\right)\
\left(1-{\beta \omega ^2 \over G_{11}}\right) ,\hspace{1.in}
\label{eq:Pvf}
\end{eqnarray}
where the limit $n \rightarrow 0$ has been taken.
The stationary points of eq.(\ref{eq:Pvf}) in the $\sigma-\mu_1$
plane as functions of $\omega $, have been obtained numerically for
various values of the parameters of the model ($\beta, \mu, g$),
and the results have been used back in eq. (\ref{eq:Pvf}) to evaluate
$P_v(\omega )$.
The results will be summarized below. But first let us examine the
simpler behavior in the tail of the probability distribution,
that can be investigated analytically.

For large values of $\omega$, the error-function in eq.(\ref{eq:Pvf})
can be approximated by
sign$(\omega)$ and the exponential term in the expression for $g(\omega )$
is negligible. If we further define the tail region as the region for which
\begin{eqnarray}
|\omega | >> \beta g /\mu \ ,
\label{eq:deftail}
\end{eqnarray}
we find that the extremum of $P_v(\omega )$ is obtained for
\begin{eqnarray}
{\sigma \over \mu _1} \approx {\beta g \over \mu |\omega|-\beta g}
\hspace{1in} \nonumber \\
\mu _1 \approx [\mu \pi + 2 \beta^3 g^2 -2 g \beta (\beta ^4 g^2
+\beta \mu \pi )^{1/2}]/\pi\ .
\label{eq:sigmu}
\end{eqnarray}
Plugging these expressions back into the expression for $P_v(\omega )$
one finds that the behavior of the spatial probability distribution
in the tail region defined by eq. (\ref{eq:deftail}) is
\begin{eqnarray}
P_v (\omega )\approx \exp(-\mu \beta \omega ^2 /2 + g \beta ^2 |\omega |
+C) \hspace{1in} \nonumber \\
C = -{\beta^3 g^2 \over 2 \mu} -2 \beta g \left({\beta \over \pi \mu_1}
\right)^{1/2} +{\mu \over 2 \mu_1} -{1 \over 2} + {1 \over 2} \ln
\left({\beta \mu_1 \over 2 \pi}\right),
\label{eq:tail}
\end{eqnarray}
where $\mu_1$ is given by eq. (\ref{eq:sigmu}) .

The extremum in the tail turns out
to be a minimum in the $\sigma -\mu _1$ plane, as opposed to the
situation for $n \geq 1$ when it is a maximum as can be seen from
eq.(\ref{eq:ineq2}). (On the contrary, for very small values of
$\omega $ we have found
numerically, that the stationarity point is actually a saddle point.)
The fact that the extremum in the tail region is a minimum seems to suggest
that the approximate expression eq.(\ref{eq:tail}) constitutes
an upper bound to the exact behavior.
The behavior in the tail given by eq.(\ref{eq:tail}) should be
compared with the behavior obtained from the conventional variational
approximation given
by substituting $\gamma = 1/2$ in eq. (\ref{eq:Phexp}). Those formulas
give a higher value for the tail than that predicted by
eq.(\ref{eq:tail}). (This is true even when $\hat{T}<1$ with the RSB
solution). More on this in the discussion below.

\section{ {\bf Comparison with numerical simulations}}
We proceed to a numerical comparison between simulations and the
results of the various variational approximations. Again, we limit the
discussion to the case $\gamma=1/2$.
\, We have studied numerically a lattice version of the toy model. \,
A suitable interval of the particle's position $\omega$
( $-12.5 /\sqrt{\beta}<\omega  < 12.5 / \sqrt{\beta}\ $) is divided
into $5,000$
lattice sites.
\, For a given realization, the algorithm generates an independently
distributed gaussian
random number for each site $r_i$; it then generates $V_j$, the
correlated random potential at site $j$, by summing the random
numbers in the following way:
\begin{equation}
V_j \, \propto\, \sum_i \, {\rm sign}(i-j) \, r_i .
\end{equation}
The quadratic term of the hamiltonian is added to this random
potential, and then the partition function and probability
distribution are calculated.

We consider three sets of values for the parameters $(\beta,
g,\mu )$ :\ \ $(0.2,2 \sqrt{\pi },1)$,\ \ $(1.0, 2 \sqrt{\pi },1)$,\ \
$(10.0,2.2,4.6)$. The corresponding values of the reduced temperature
$\hat{T}$ for these three cases are $5$, $1$ and $0.23$ respectively.
The data for $\hat{T}=5$ is plotted in Fig.1 (solid curve). It was
obtained from numerical simulations with 50000 realizations of the
disorder. The dashed line represents the probability $P_h$ given in
eq.(\ref{eq:Phexp}) and derived from the hamiltonian variational
method. The diamonds represent the approximation developed in the last
section. We see that it gives a perfect fit to the data for this
value of $\hat{T}$. We have used Mathematica to find the stationary
point of eq.(\ref{eq:Pvf}) in the $\mu_1-\sigma$ plane and then
substituted these values back into eq.(\ref{eq:Pvf}).
The asymptotic formula eq.(\ref{eq:tail}) predicts
\begin{eqnarray}
P_v(\omega)\approx \exp(-0.1\omega ^2+0.142\omega -2.1)
\end{eqnarray}
where we used $\mu_1=0.7$ from eq. (\ref{eq:sigmu}). This behavior is
also indistinguishable from the data in the range $10<\omega<28$.

Let us proceed to the case $\hat{T}=1$. The data is plotted in Fig.2.
In this case the noise associated with the random potential is apparent.
In this case we found it necessary to average over 500000
realizations. The data for 50000 realizations is also shown in lighter
dots. In the tail region it is apparent that the curve corresponding
to the lower number of realizations has a lower value, since the
average is increased by relatively rare events. The dashed curve is
the single parameter hamiltonian variational fit. The new variational
method results are represented by the diamonds. The asymptotic
formulas predicts $\mu_1=0.06$ and
\begin{eqnarray}
P_v(\omega)\approx \exp(-0.5\omega ^2+3.545\omega -17).
\label{eq:asym02}
\end{eqnarray}
It is represented by the solid line in the figure.
The diamonds do not lie exactly on this line, because $\omega$ is
still not large enough in this range. Again we see that the
variational approximation gives an excellent fit to the data.

The final example is for $\hat{T}=0.23$ using $10^6$ realizations of
the disorder. The data is depicted in Fig. 3. The dashed line is the
result of the hamiltonian RS variation. The dot-dashed line is the
result of the RSB hamiltonian variation which is the appropriate
solution for
$\hat{T}<1$. The diamonds again represent our new variational method
results. The range of $\omega$
simulated is lower than the onset of the tail region given by
eq.(\ref{eq:deftail}) which in the present case starts above
$\omega\sim 5$, so a comparison with the asymptotic
formula eq.(\ref{eq:tail}) is not shown.

We see that the data in this case falls below the result of our variational
method, which is nonetheless better than the hamiltonian RSB variational
approximation. Two possibilities come to mind to explain this
discrepancy.

1. It seems quite possible that because the relative strength of the random
part of the potential is much larger
when $\hat{T}$ is small, as compared to the harmonic part, one needs
more realizations to accumulate enough statistics for the averaged
probability
distribution. For $\hat{T}=5$ even $5000$ realizations
already gave us good results. In Fig.1 we show the results for 50000
realizations, but these are practically indistinguishable from the
average of
5000 realizations. For $\hat{T}=1$, 50000 realizations are not
sufficient and we had
to average at least 500000 realizations to accumulate enough statistics.
This is because we have to collect enough rare events which contribute to
the distinction between the average and typical values of the spatial
probability distribution. \cite{gb1,gb2}
It is quite plausible that to get better results for $\hat{T}=0.23$
one has to go to a higher number of realizations. Averaging over more
realizations (see e.g Fig. 2 for the distinction between 50000 and 500000
realizations) may narrow the gap between the data and the
variational approximation.

2. Another strong possibility is that like the hamiltonian variational
case it is necessary for small $\hat{T}$ to look for a solution with replica
symmetry breaking when using the current variational scheme. Such a
solution may display a different asymptotic behavior for small $\hat{T}$
than the one given by eq.(\ref{eq:tail}). It may also yield a lower
value in the pretail region (which is the range of values depicted in
Fig. 3). In order to find such a
solution one has to go back to the full expressions derived in Sec. 2
without making the simplifying assumptions of replica symmetry made in
Sec. 3, and one has to extremize the probability distributions under
the more general conditions.
The task of finding of a RSB solution is left for future research.

\section{ {\bf Discussion}}
In this paper we have developed a new variational scheme for
approximating the spatial probability distribution. The replica
symmetric solution gives an
excellent approximation to the numerical data for high temperatures
$\hat{T}>1$. For the tail of the spatial probability it predicts a
gaussian decay with a linear exponential correction. At lower
temperature, the fit in tail region is not that good, either because
the numerical data is insufficient and one needs to average over a
higher number of realizations, or a solution with replica symmetry
breaking is needed (or both). Of course since we are dealing after all
with an approximation, we are never guaranteed a perfect fit. For all
temperatures tested the present method gives a much better fit to the
probability distribution than using the hamiltonian variational
method, both with or without RSB.

When the data for $\hat{T}=0.23$ is displayed in a log-log plot, \cite{gb2}
a crossover is observed from a behavior $P\ \sim \exp(-{\rm const}\
\omega^2)$ to a behavior like $P\ \sim \exp (-{\rm const}\ \omega^3)$.
An empirical fit to the data of the form
\begin{eqnarray}
P(\omega) \sim \exp (-3.5 \omega^3 +20 )
\end{eqnarray}
is depicted as open triangles in Fig. 3 for the range $ 6 <\omega^2 < 16$.
Since in this case we are not really observing the tail region
(according to the definition given in eq. (\ref{eq:deftail}) ), this
behavior may be a transient or intermediate behavior. We should also
be cautious about the data in this region, because as mentioned in the
last section it is possible that more realizations are needed.

On the other hand, if we do take the cubic $\omega$ dependence of the
log of the probability distribution seriously, this immediately rings a bell
because of some work done by Villain {\it et al.} \cite{villain} on
the toy model.
What they actually showed was that the probability for rare realizations
of the disorder which make $H(\omega) \sim 0$ and thus $\exp(-\beta H
)\sim 1$
behaves like $\sim \exp (- \omega^3 \mu^2 /2 g)$, for values of
$\omega^3 >> 2 g/\mu^2$ (I have transformed their notation to ours).
This could lead to the conclusion that the
spatial probability distribution should also behave like a cubic power
of $\omega$. This is true only if the probability distribution is
completely dominated by the rare realizations of this kind, a fact
that was never claimed by Villain {\it et al.}. It is certainly possible
that other realizations which give lower values to $\exp(-\beta H)$
but are nonetheless more abundant dominate the average value.
Villain's argument is valid for any temperature. Our numerical
results show that one certainly does not get a cubic behavior of the
probability distribution at high temperature for values of $\omega >>
(g/\mu^2)^{1/3}$. In that case the rare realizations of of the type
considered by Villain still exist but they do not dominate the
probability distribution.

This is actually easy to understand. Let us define two different
length scales in the problem. The first, introduced by Villain {\it et
  al.} which we call $\xi_1$ is defined as
\begin{eqnarray}
\xi_1 \simeq \left(2 g \over \mu^2 \right)^{1/3}.
\label{eq:xi1}
\end{eqnarray}
is the length above which the probability for rare events of magnitude
1 goes like
\begin{eqnarray}
A \exp(-{\omega^3 \over \xi_1^3})
\label{eq:o3}
\end{eqnarray}
with some undetermined constant A.
The second length, which I introduced in eq.(\ref{eq:deftail}),
\begin{eqnarray}
\xi_2 \simeq {\beta g \over \mu}\ ,
\label{eq:xi2}
\end{eqnarray}
is the length for which the asymptotic behavior in the tail starts
according to our investigation in Sec. 3. For $\omega$ above this
value we have found the behavior
\begin{eqnarray}
P(\omega) \simeq \exp(-\beta \mu \omega^2 /2 +g \beta^2 |\omega|
+C(\beta))\ ,
\label{eq:o2}
\end{eqnarray}
where $C(\beta)$ is given in eq. (\ref{eq:tail}). Since
\begin{eqnarray}
\xi_2 \simeq \xi_1 \times \ {(2 \pi)^{1/3} \over \hat{T}}
\label{eq:xi1xi2}
\end{eqnarray}
We see that for $\hat{T} > 1.8 $, the cubic behavior is never
realized since the contribution of the rare realizations of the
Villain type to the averaged probability is smaller than the behavior
given by eq.(\ref{eq:o2}). This explain the perfect fit of the
asymptotic behavior given in eq.(\ref{eq:asym02}) and the numerical
data, starting at a very low value of $\omega$.

On the other hand for $\hat{T} << 1.8 $ it is conceivable to have two
tail regimes,
the first with $\xi_1 < \omega < \xi_2$ in which the probability
has the cubic behavior in $\omega$ because it is dominated by rare
configurations of the Villain type, and a second tail regime
for $ \omega > \xi_2$ for which the asymptotic behavior is changed
to a quadratic behavior given by eq.(\ref{eq:o2}), if the results of
the RS solution are valid (or possibly
to a different asymptotic behavior which will emerge from a RSB solution).
This is because one can easily check that at the border between these
two regimes the two
expressions given by eq.(\ref{eq:o3}) and eq.(\ref{eq:o2}) become comparable
in magnitude, and the quadratic behavior wins for $\omega > \xi_2$.
As the (reduced) temperature is lowered from  1 the first regime (cubic
behavior) is expected to grow in size and include all of the tail at $\hat{T}
=0$. Simulations of directed polymers at zero temperature, which can
be mapped into corresponding results for the toy-model,
\cite{halpin} showed the beginning
of a change in the form of the spatial probability distribution from
$\sim \exp (- c \omega^2)$ to $\sim \exp (- c \omega^{2.2})$ at the onset
of the tail,
but did not go far enough into the tail region to confirm a cubic dependence.
See also ref.\cite{zumophen}.
We have previously observed cubic dependence in the tail of the
directed polymers' spatial probability distribution at finite
temperature. \cite{gb1}

There is some difficulty though, to explain the apparent cubic behavior in
Fig. 3 purely in terms of Villain's configurations because we
simulated ``only'' over $10^6$ realizations and thus if the average is
dominated by a single realization whose contribution is unity, the
value of the average should be of the order of $10^{-6}\simeq
\exp(-14)$ which far exceeds the value of the distribution in most of
this region (except at the very beginning). One might
argue that there are other realizations with somewhat smaller
contributions to the average than Villain's which also gives rise to a
similar cubic behavior but this needs to be verified.

We hope that this work will stimulate further investigation of the
behavior of the tail of the probability distribution at low
temperatures. Three important questions that need a definite answer are

\noindent 1. Is there a RSB solution to our new variational equations?

\noindent 2. Is a cubic dependence of the log of the probability
indeed realized over a large region when the temperature is very low?

\noindent 3. Is the asymptotic behavior given by eq.(\ref{eq:tail})
which works so well for high temperatures when $\omega>\xi_2$, also
valid asymptotically at low temperatures?

We also hope that it will be possible to extend the variational method
developed in this paper for the zero-dimensional manifold directly to higher
dimensional manifolds in random media.
\vspace{0.1in}

\noindent {\bf Acknowledgements}

I thank Professor G. Parisi for instigating the current investigation,
and for his kind hospitality at the University of Rome, where this
research has begun.
Support of the National Science Foundation under grant number DMR-9016907
is gratefully acknowledged. We also acknowledge Cray time allocation
from the Pittsburgh Supercomputer center.

\newpage
\appendix
\section{}

In this Appendix we derive eq.(\ref{eq:avHmh}) starting from eq.
(\ref{eq:trHmh}).
Let us shift the variables $\omega_a$ in eq.(\ref{eq:trHmh})
\begin{eqnarray}
\omega_a \rightarrow \omega_a +\lambda_a
\label{eq:shift}
\end{eqnarray}
such that the linear term in $\omega _1$ in the exponential
is eliminated. This is achieved by choosing
\begin{eqnarray}
\lambda _a=-(i/\beta ) G_{1a} k
\end{eqnarray}
We then find:
\begin{eqnarray}
\widehat{Tr} \ [({\cal  H}\ -\ h\ )\ e^{\ -\beta h}\ ] = \int {dk
  \over 2 \pi} e^{ik\omega}\ e^{-{1 \over 2 \beta }G_{11}k^2}
\int d\omega_1 \cdots d\omega_n  \ \ \ \ \ \ \ \ \  \ \ \ \ \ \ \ \ \
\ \ \ \ \ \ \ \ \ \ \ \ \ \ \ \ \ \ \  \nonumber  \\
\times \left({\beta \over 2}  \sum_{ab}
f((\omega_a-\omega_b+\lambda_a -\lambda_b )^2) + {1 \over 2}
\sum_{ab}\sigma_{ab} (\omega_a+\lambda_a)(  \omega_b+\lambda_b )
 \right) \nonumber \\
\times \exp \left( -{\beta   \over 2} \sum_{ab} (G^{-1})_{ab}
\omega_a\omega_b \right) \ .\hspace{2.35in}
\label{eq:trHmh2}
\end{eqnarray}
We now expand the function f in a power series about $0$, (this is used
only as a tool derive our result. It may not be necessary for f to be
analytic about 0)
\begin{eqnarray}
f(x)=\sum_{l=0}^\infty f_l \ x^l
\end{eqnarray}
We use the following formula to integrate over $\omega_1 \cdots  \omega _n$:
\begin{eqnarray}
\int d\omega_1 \cdots d\omega_n\ (\omega_a -\omega_b )^{2 s}
\  \exp \left( -{\beta   \over 2} \sum_{ab} (G^{-1})_{ab}
\omega_a\omega_b \right)= \nonumber \\
(2 \pi / \beta )^{n/2} ({\rm Det}\ G)^{1/2}
{2^s \Gamma(1/2+s) \over \beta ^s \Gamma(1/2)} (G_{aa}+G_{bb}-2G_{ab})
^s \ ,\ \
\label{intomega}
\end{eqnarray}
provided s is a non-negative integer. For $s$ being half-integer
the integral is $0$.
We obtain
\begin{eqnarray}
{\widehat{Tr} \ [({\cal  H}\ -\ h\ )\ e^{\ -\beta h}\ ]
\over \widehat{Tr} \ \ e^{\ -\beta h}\ }
= \left({G_{11} \over 2 \pi \beta}\right)^{1/2}
\exp \left( {\beta \omega^2 \over 2 G_{11}}\right)
\int dk \left\{ \ \exp \left({ik\omega\ -{1 \over 2 \beta
    }G_{11}k^2}\right) \right. \times \hspace{0.2in} \nonumber  \\
\left. {\beta \over 2}  \sum_{ab}
\hat{f}\left(-(G_{1a}-G_{1b})^2 k^2/\beta^2,\ (G_{aa}+G_{bb}-2 G_{ab})
/\beta \ \right)
 \exp \left( -{\beta \over 2} \sum_{cd} (G^{-1})_{cd} \omega_c\omega_d
\right) \right\}  \nonumber \\
-{n \over 2 \beta }+{\mu \over 2 \beta} \sum_a G_{aa}
+{1 \over 2 \beta G_{11}} \left(G_{11}-\mu \sum_a G_{1a}^2\right)
\left(1-{\beta \omega^2\over  G_{11}}\right) \ ,\hspace{1.75in}
\label{eq:tr3}
\end{eqnarray}
where
\begin{eqnarray}
\hat{f}(x,y)={1 \over \sqrt{\pi}}\sum_{l=0}^\infty\ f_l
\ (2y)^l\ \sum_{s=0}^l g(l,s) \Gamma(s+1/2) \left({x\over 2y}
\right)^{l-s} \nonumber \\
g(l,s)=\sum_{j=0}^l\sum_{m=0}^j \ \delta_{(j+m)/2,s}
\left( {l \atop j}\right)\left( {j \atop m}\right)2^{j-m}.
\hspace{0.74in}
\end{eqnarray}
Since
\begin{eqnarray}
\int_{-\infty}^{\infty} dp \ e^{-p^2} (p+q)^{2l}=\sum_{s=0}^l
g(l,s)\Gamma (s+1/2) q^{2(l-s)},
\end{eqnarray}
we see that $\hat{f}(x,y)$ can be expressed in the form
\begin{eqnarray}
\hat{f}(x,y)={1 \over \sqrt{\pi}}\int_{-\infty}^{\infty}
 dp \ e^{-p^2}\ f\left(2y\left(
p+\sqrt{x \over 2y}\right)^2\right) \hspace{.1in}\nonumber \\
=e^{-{x \over 2y}} {1 \over \sqrt{\pi}}\int_{-\infty}^{\infty}
dp \ f(2yp)\ e^{-p^2+2\sqrt{x/(2y)}\ p} \ .
\end{eqnarray}
Using this representation for $\hat{f}(x,y)$ the integral over $k$
in eq.(\ref{eq:tr3}) can be performed, and we obtain the desired
result given in eq.(\ref{eq:avHmh}).

\newpage
{\bf Figure Captions}

Figure 1: Plot of the log of spatial probability distribution vs.
$\omega^2$ for $\hat{T}=5$ (solid line). The dashed line is the result of the
hamiltonian variational approximation and the diamonds the results of
our new variational scheme.

Figure 2: Plot of the log of the spatial probability distribution vs.
$\omega^2$ for $\hat{T}=1$. The wiggly solid curve represent the
result of averaging over 500000 realizations. The light wiggly curve
is for 50000 realizations. The dashed curve and the diamonds are
explained in the caption of Fig. 1. The solid smooth curve result from
the asymptotic formula, eq.(\ref{eq:tail}).

Figure 3: Plot of the log of the spatial probability distribution vs.
$\omega^2$ for $\hat{T}=0.23$. The wiggly solid curve represent the
result of averaging over $10^6$ realizations. The dashed and
dashed-dotted lines represent the hamiltonian RS and RSB approximations
respectively. The diamonds are the results of the new variational scheme.
The open triangles represent the empirical cubic approximation.

\end{document}